\documentstyle[prb,aps]{revtex}
\begin{document}
\input{psfig}
\twocolumn[\hsize\textwidth\columnwidth\hsize\csname @twocolumnfalse\endcsname
\title{Protein structures and optimal folding emerging from a
geometrical variational principle}
\author{Cristian Micheletti$^1$, Jayanth R. Banavar$^2$, Amos
Maritan$^1$ and Flavio Seno $^3$}
\address{(1) INFM and International School for Advanced Studies (S.I.S.S.A.),
Via Beirut 2-4, 34014 Trieste, Italy\\
The Abdus Salam Centre for Theoretical Physics - Trieste, Italy}
\address{(2) Department of Physics and Center for Materials Physics,
104 Davey Laboratory, The Pennsylvania State University, University
Park, Pennsylvania 16802}
\address{(3) INFM - Dipartimento di Fisica, Via Marzolo, 8 , 35100
Padova, Italy} 
\date{\today}
\maketitle
\begin{abstract}
Novel numerical techniques, validated by an analysis of barnase and
chymotrypsin inhibitor, are used to elucidate the paramount role
played by the geometry of the protein backbone in steering the folding
to the correct native state.  It is found that, irrespective of the
sequence, the native state of a protein has exceedingly large number
of conformations with a given amount of structural overlap compared to
other compact artificial backbones; moreover the conformational
entropies of unrelated proteins of the same length are nearly equal at
any given stage of folding.  These results are suggestive of an
extremality principle underlying protein evolution, which, in turn, is
shown to be associated with the emergence of secondary structures.
\end{abstract}
]

The rapid and reversible folding of protein-like heteropolymers into
their thermodynamically stable native state [\onlinecite{Anf}] is
accompanied by a huge reduction in conformational entropy
[\onlinecite{KarplusW,Pt92}].  Evidence has been accumulating for an
achievement of the entropy reduction through a folding funnel which
favors the kinetic accessibility of the native state
[\onlinecite{due,uno,LMO,tre,trebis}].  Some fundamental questions
remain, however, unanswered.  What makes proteins special compared to
random heteropolymers?  What guides the folding of a protein?  Is it
the sequence that is fundamental or its native structure?

In this letter, we examine these issues and focus on the special role
played by the native structure of proteins, with no input of
information regarding amino acid sequences.  The study is carried out
through a novel theoretical probe for the conformation space of
proteins: a measure of the density of alternative conformations (DAC)
having a given overlap or percentage of contacts in common with a
fixed native structure. We demonstrate with studies on chymotrypsin
inhibitor (2ci2) and barnase (1a2p) that the DAC provides key
information on the folding nucleus [\onlinecite{F95}].  An analysis of
the DAC for real protein structures and for artificially generated
decoy ones suggests that an extremal principle is operational in
nature, which maximizes the DAC at intermediate overlap, providing a
large basin of attraction [\onlinecite{DC,BOSW95,uno,LMO,tre}] for the
native state and promoting the emergence of secondary structures.

Operationally, our study consists of the determination of the number
of alternative structures which have a given structural similarity to
a putative native state.  The structural similarity between the native
structure and an alternative one is defined as the percentage of
common native contacts in the alternative conformation.  It is well
known that such a measure is a good coordinate characterizing the
folding process [\onlinecite{go,CT,Sali94b}].  Following standard
practice, two residues are defined to be in contact if the distance
between their $C_\alpha$ atoms is less than $6.5 \AA$.  In an unbiased
study, conformations that differ slightly should not be considered
distinct.  To avoid this problem, we perform a coarse-graining of the
configurational degrees of freedom by adopting the discretization
approach introduced by Covell and Jernigan [\onlinecite{CJ}], where
the $C_\alpha$'s occupy sites on a suitably oriented FCC lattice (of
edge 3.8 $\AA$).  This discretization does not distort the peptide
angles and the position of the coarse-grained $C_\alpha$'s differ from
the true ones by typically less than 1 $\AA$ RMSD [\onlinecite{PL95}].
For proteins of about 100 residues, the contact maps
[\onlinecite{Levit}] of the real and FCC coarse-grained contacts maps
are virtually identical.

The generation of alternative conformations was carried out using a
Monte Carlo procedure. A starting conformation was successively
modified by displacing the $C_\alpha$'s to unoccupied positions of the
FCC lattice. The move of an amino acid to an unoccupied site is
allowed only if the new conformation satisfies certain constraints of
steric overlap and peptide geometry.  These constraints (any two
non-consecutive residues cannot be closer than $4.65 \AA$ due to
excluded volume effects and the peptide bond is not stretched beyond
$5.37 \AA$) were determined after carrying out an FCC coarse-graining
of several proteins of intermediate length ($\approx 100$ residues)
and enforced in the generation of alternative protein-like
conformations.

In order to minimize the effects of correlation between successively
generated structures, we typically discarded 50 elementary moves
before accepting each new conformation.  A newly generated
conformation was accepted with the usual Metropolis rule according to
the change in the Boltzmann weight: $e^{\Delta / K_B T}$, where
$\Delta$ is the change in contact overlap and $T$ is a fictitious
temperature. By choosing $T$ appropriately, one can readily generate
alternative conformations with a desired average contact overlap,
$\bar{q}$. At a given temperature, the true number of alternative
structures with overlap $q$ is proportional to the number of states
with overlap $q$ obtained in the simulation multiplied by the
Boltzmann weight. On undoing the Boltzmann bias, it is possible to
recover the true density of states in a region around $\bar{q}$. In
order to obtain the density of states for all values of overlap, we
performed Monte Carlo samplings at different temperatures and then
used standard deconvolution procedures [~\onlinecite{FS}].

\begin{figure}
\centerline{\psfig{figure=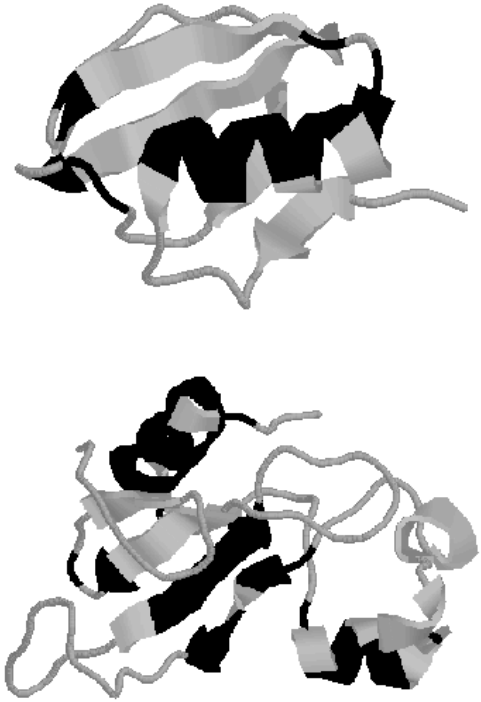,width=3.1in}}
\caption{Ribbon plot (obtained with RASMOL) of 2ci2 (top) and barnase
(bottom). The residues involved in the 12 [16] most frequent contacts
of alternative structures with overlap $\approx 40 \%$ with the native
conformations are highlighted in black. The majority of these coincide
with contacts that are formed at the early stages of folding.}
\label{fig:2ci2}
\end{figure}

\begin{figure}
\centerline{\psfig{figure=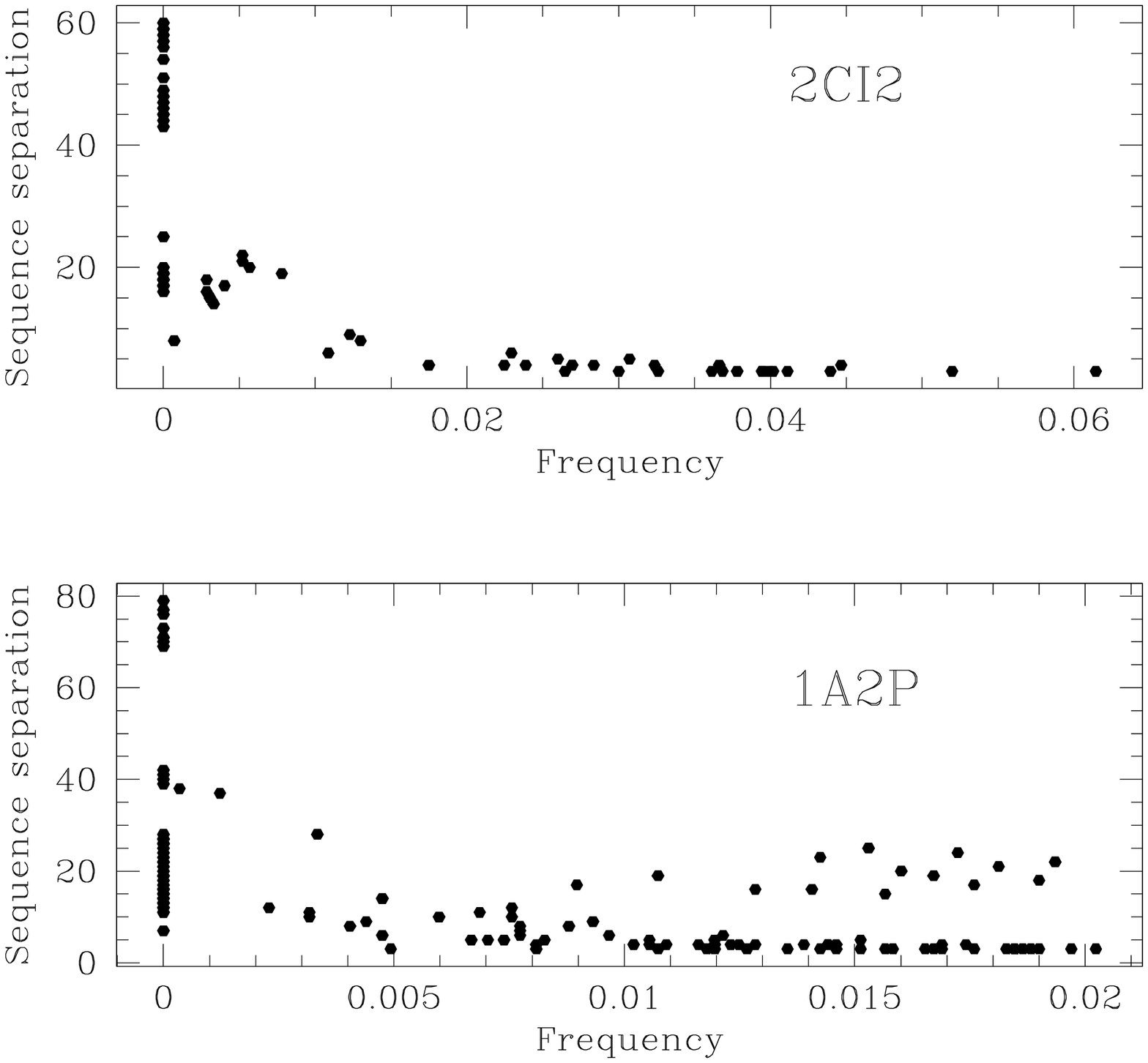,width=3.1in}}
\caption{Distribution of sequence separation of contacts common in
alternative conformations for
2ci2 and 1a2p.  The most frequent contacts in 2ci2 have
a small sequence separation (3-4) and pertain to helix formation.
1a2p shows a very different
behaviour with several contacts with very large sequence separation. }
\label{fig:fsep}
\end{figure}

We begin with the backbones of the chymotrypsin inhibitor (2ci2) and
barnase (1a2p) and generated alternative structures with a not too
large overlap [\onlinecite{footnote1}] ($\approx 40 \%$) for each of
them. It turned out that the most frequent contacts shared by the
native conformation of 2ci2 with the alternative ones involved the
helical-residues 30-42 (see top Fig. \ref{fig:2ci2}) and the rarest
ones pertained to interaction between the helix and $\beta$-strands
and between the $\beta$-strands themselves.  This is in excellent
agreement with the studies of Fersht {\em et al.}
[\onlinecite{F95,ci2b}], which demonstrated the formation of the helix
at early stages of the folding. A different behaviour (see bottom
Fig. \ref{fig:fsep}) was found for barnase, where, again, for overlap
of $\approx 40 \%$, we find many contacts pertaining to the nearly
complete formation of helix 1 (residues 8-18), a partial formation of
helix 2, in particular bonds between residues 26-29 and 29-32 as well
as several non-local contacts bridging the $\beta$-strands, especially
residues 51-55 and 72-75. This picture is fully consistent with the
experimental results obtained in ref. \onlinecite{barnase}.

This provides a sound {\em a posteriori}\/ justification that the main
features of the folding of a protein can be followed from a study of
the DAC. Remarkably, the method discussed above relies entirely on
structure-related properties and suggests that the features of the
folding funnel are determined by the geometry of the ``bare''
backbone, while the finer details, of course, depend on the specific
well-designed sequence.

We now turn to an analysis of three proteins of length 51 (1hcg, 1hja
and 1sgp) which have nearly the same number of native contacts
($\approx 83$). For each structure, we calculated the DAC with the
constraint that the total number of contacts in the alternative
structures do not exceed 88 to avoid excessive compactness.  In order
to assess whether the DAC associated with naturally occurring proteins
had special features, we generated three decoy compact conformations
of the same length and number of contacts, but with different degrees
of short and long range contacts (in sequence separation).  These
decoys (subject to the aforementioned ``physical constraints'') were
generated with a simulated annealing procedure to find the structure
with the highest overlap with a target contact matrix.  By tuning the
number of short-range versus long-range entries in the target random
contact matrix, we generated three structures with different degree of
compactness and local geometrical regularity.

\begin{figure}
\centerline{\psfig{figure=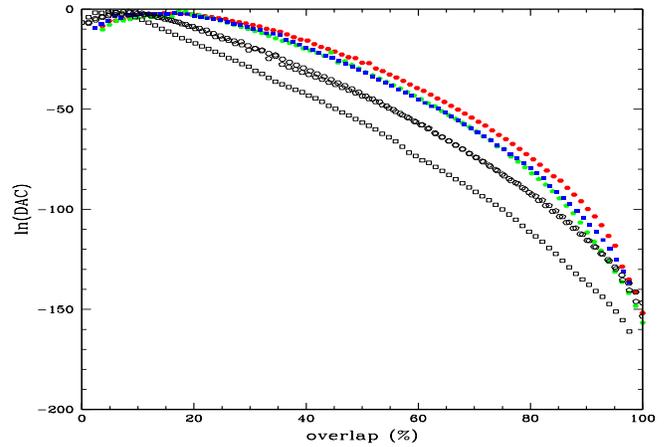,height=2.5in,width=3.5in}}
\caption{Density of states for proteins for 1sgp (filled squares),
1hja (filled pentagons) and 1hcg (filled hexagons). Curves for
artificial decoy structures are denoted by the open symbols.}
\label{fig:dos1}
\end{figure}

The plots of the DAC are shown in Fig. \ref{fig:dos1}.  A striking
feature of the curves is that, for intermediate overlap, the DAC of
the real proteins is enormously larger than that of the decoys (note
the logarithmic scale) and suggests that naturally occuring
conformations have a much larger number of entryway structures than
random compact conformations.  Furthermore, for very high values of
the overlap, the steepness of the protein curves is much larger than
those of the decoys, showing that the reduction in the conformational
entropy is also correspondingly higher.  This translates into the
existence of a funnel with a very large basin and steep walls.
Another significant feature is the good collapse of the protein
curves.  We have verified that this feature also obtains for 1bd0 and
2pk4 which each have 80 residues and 140 and 146 contacts
respectively.  A simple explanation for the curve collapse could be
that the density of states for real proteins is ``extremal'', in that
it is close to the maximum possible value for intermediate values of
the overlap.

The importance of the locality of contacts for folding kinetics was
highlighted recently by Plaxco {\em et al.} [\onlinecite{Pl}] who
found a correlation between folding rate and contact order, defined as
the average sequence separation of contacts normalized to the total
number of contacts and sequence length.  With reference to
Fig. \ref{fig:dos1}, the contact order value for protein 1hcg, 1 hja
and 1sgp is 0.139, 0.214 and 0.204 respectively. For the decoy
structures, it is 0.424, 0.222 and 0.179 for the curves denoted by
open squares, pentagons and hexagons, respectively.  
The lowest curve in the figure is indeed associated
with an unusually high contact order in accord with the findings of
Plaxco {\em et al.} [\onlinecite{Pl}].

A ubiquitious feature of protein structures is the existence of
secondary structure motifs [\onlinecite{Cra,Li}].  We have carried out
some simple investigations to assess whether a correlation exists
between the extremality of the DAC curve and the emergence of
secondary-structure-like motifs.

We considered a space of contact maps [\onlinecite{Levit}], within
which each of the residues interacted with the same number of other
residues, $n_c$ (typically $n_c=5$, as in the average case of a
protein with about 100 residues and a cutoff distance of 6.5 $\AA$).
This space contains both maps corresponding to real structures and
unphysical ones.  Furthermore, to mimic the effects of the rigidity
and geometry of the peptide bond, we disallowed contacts between
residue $i$ and the four neighboring residues along the sequence
$i-2$, $i-1$, $i+1$ and $i+2$.

 In this context, the maximization of the density of states
corresponds to finding the target matrix with the highest number of
matrices sharing a given fraction of its contacts.  Although it is
difficult to solve this problem, for arbitrary values of the overlap,
it is relatively easy to generate matrices with an overlap close to
the maximum value, $\bar{q}_{max}$ (for a $L$x$L$ matrix,
$\bar{q}_{{max}}=L\cdot n_c$).  To enumerate all matrices with overlap
$\bar{q}_{{max}}-2$, one first identifies a pair of non-zero entries
in the target matrix ${\bar m}$: $\bar{m}_{ij}=\bar{m}_{kl}=1$. Then
it is necessary to check whether entries $\bar{m}_{il},\bar{m}_{kj}$
are both ``free'' (i.e. equal to zero) and do not correspond to
forbidden contacts (e.g. between $i$ and $i+1$). If this is so, the
old pair of entries (and their symmetric counterpart) are set to zero,
and the new ones to 1.  By considering, in turn, all possible pairs of
non-zero entries one can generate all matrices of overlap
$\bar{q}_{{max}}-2$ . Then, by performing a simulated annealing in
contact-map space one can isolate the map having the highest number of
matrices with overlap $\bar{q}_{max}-2$.

\begin{figure}
\centerline{\psfig{figure=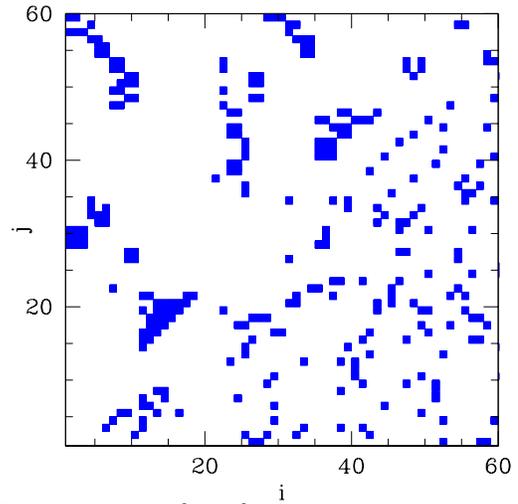,width=2.7in}}
\caption{The upper [lower] triangle shows a target contact matrix with
$L=60$ that has a large [intermediate] number of contact maps with an
overlap of $\bar{q}_{{max}}-2$ contacts.}
\label{fig:map}
\end{figure}

We carried out our calculations for values of $L$ around 60. The
optimal matrices appear to have features reminiscent of
$\alpha$-helices and $\beta$-sheets, as shown in Fig. \ref{fig:map}.
A more quantitative measurement of the secondary-structure content of
the optimal matrices can be obtained by considering the correlation
functions

\begin{equation}
g_1 (x) = \sum_{i}  m_{i,i+x} \ ; \ \ \ \ g_2 (x) = \sum_i  m_{i,x-i}
\label{eqn:corr}
\end{equation}

\noindent which show peaks in correspondence with the sequence
separation of residues involved in $\alpha$-helices and parallel
$\beta$-sheets ($g_1$) or antiparallel $\beta$-sheets
($g_2$).

\begin{figure}
\centerline{\psfig{figure=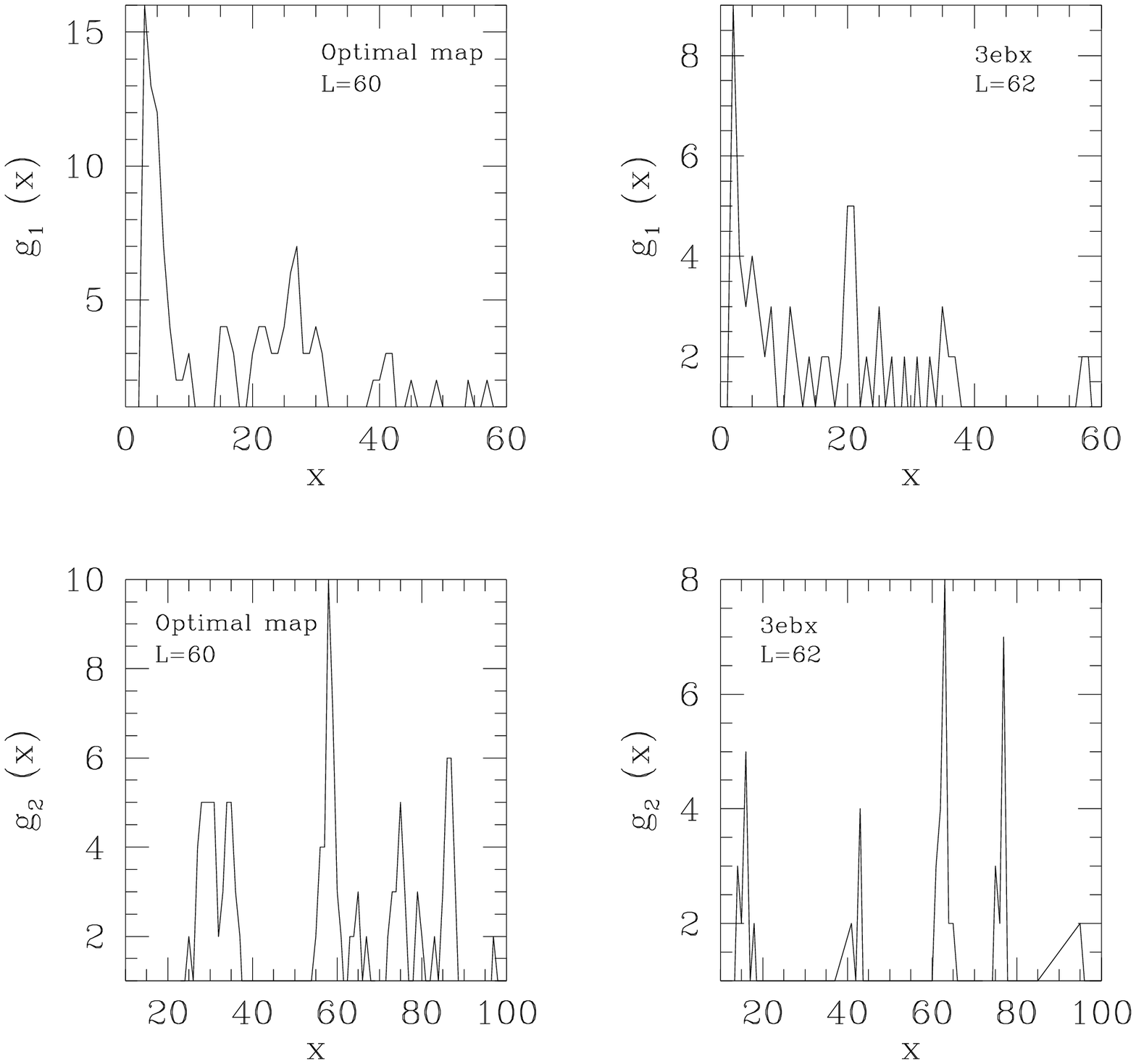,width=3.0in}}
\caption{Correlation functions (see equation 1) for an
optimal target matrix of length 60 and for protein 3ebx.}
\label{fig:corr}
\end{figure}

A typical plot of the correlation functions for an optimal map of
length 60 and for the protein 3ebx (length 62) are shown in
Fig. \ref{fig:corr}.  The similarity of the plots is striking,
particularly because, in both cases, the height of the peaks in $g_1$
decreases with sequence separation, unlike the situation with $g_2$.

In summary, novel numerical techniques are used to elucidate the
paramount role played by the geometry of the protein backbone in
providing a large basin of attraction to the native state. It is found
that, irrespective of the sequence, the native state of a protein has
an exceedingly large number of conformations with a given amount of
structural overlap compared to other compact artificial backbones.
Strikingly, by studying the conformational entropy of a backbone it is
possible to identify the folding nucleus with no input of the actual
protein sequence.  Moreover, the conformational entropies of unrelated
proteins of the same length are nearly equal at any significant value
of the reaction coordinate [\onlinecite{footnote1}]. These results are
suggestive of an extremality principle underlying the selection of
naturally occurring folds of proteins which, in turn, is shown to be
associated with the emergence of secondary structures. Our procedure
ought to be useful for the generation of alternative conformations
necessary for protein design and the determination of the effective
interactions between amino acids.

ACKNOWLEDGEMENTS. We acknowledge support from INFM, NASA, NATO.

\begin{references}
\bibitem{footnote1} The significance of the overlap is measured
against the typical overlap of any two compact-like conformations,
which is about 10-20 \% (consistent with the unfolding simulations of
ref. [\onlinecite{karp}]).
\end{references}
\end{document}